\documentclass[aps,prl,reprint,twocolumn,superscriptaddress]{revtex4-1}

\usepackage{euscript}
\usepackage{amssymb}
\usepackage{amsfonts}
\usepackage{amsbsy}
\usepackage{epsfig}
\usepackage{amsthm}
\usepackage{amscd}
\usepackage{amstext}
\usepackage{verbatim}
\usepackage{amsmath}
\usepackage{cancel}
\usepackage{capt-of}
\usepackage{empheq}
\usepackage{subfigure}
\usepackage{xcolor}

\usepackage[pdftex]{hyperref}
\hypersetup{ 
colorlinks=true, 
linkcolor=black, 
citecolor=red, 
}

\usepackage{amssymb}

\def\ben{\begin{equation}}
\def\een{\end{equation}}

\let\l=\lambda

\let\w=\omega  \let\D=\Delta

\let\pa=\partial
\def\be{\begin{equation}}
\def\ee{\end{equation}}
\def\beq{\begin{equation}}
\def\eeq{\end{equation}}
\def\ba{\begin{array}}
\def\ea{\end{array}}

\def\dalemb#1#2{{\vbox{\hrule height .#2pt
       \hbox{\vrule width.#2pt height#1pt \kern#1pt
               \vrule width.#2pt}
       \hrule height.#2pt}}}

\newcommand{\bea}{\begin{eqnarray}}
\newcommand{\eea}{\end{eqnarray}}
\DeclareMathOperator{\tr}{tr}

\makeatletter
\newcommand*\bigcdot{\mathpalette\bigcdot@{.5}}
\newcommand*\bigcdot@[2]{\mathbin{\vcenter{\hbox{\scalebox{#2}{$\m@th#1\bullet$}}}}}
\makeatother

\renewcommand{\eqref}[1]{(\ref{#1})}

\def\ocal{{\mathcal{O}}}

\newcommand{\expect}[1]{{\langle {#1} \rangle}}
\newcommand{\mcal}{\mathcal{M}}

\begin{document}
\frenchspacing

\title{Bootstrapping Matrix Quantum Mechanics}
\author{Xizhi Han, Sean A. Hartnoll, Jorrit Kruthoff \\ {\it Department of Physics, Stanford University,} \\ {\it Stanford, CA 94305-4060, USA}}



\begin{abstract}

Large $N$ matrix quantum mechanics is central to holographic duality but not solvable in the most interesting cases. We show that the spectrum and simple expectation values in these theories can be obtained numerically via a `bootstrap' methodology.
In this approach, operator expectation values are related by symmetries --- such as time translation and $SU(N)$ gauge invariance ---
and then bounded with certain positivity constraints.
We first demonstrate how this method efficiently solves the conventional quantum anharmonic oscillator. We then reproduce the known solution of large $N$ single matrix quantum mechanics. Finally, we present new results on the ground state of large $N$ two matrix quantum mechanics.

\end{abstract}

\maketitle

{\it Introduction.---}\label{intro}
Large $N$ matrices are at the heart of the holographic emergence of semiclassical, gravitating spacetime geometry \cite{Maldacena:1997re}. In matrix quantum mechanics geometry emerges from an underlying theory with no built in locality. The simplest such theory is the single matrix quantum mechanics description of two dimensional string theory \cite{Klebanov:1991qa}, while the richest are the maximally supersymmetric multi-matrix theories of BFSS \cite{Banks:1996vh} and BMN \cite{Berenstein:2002jq}. There are many theories in between, with varying numbers of matrices and degrees of supersymmetry \cite{deWit:1997ab}. Thus far, only the single matrix quantum mechanics has proved solvable at large $N$ \cite{Brezin:1977sv}.

Nonzero temperature Monte Carlo studies of large $N$ multi-matrix quantum mechanical systems have successfully captured aspects of a known dual spacetime in supersymmetric theories \cite{Anagnostopoulos:2007fw, Catterall:2008yz, Filev:2015hia, Berkowitz:2016jlq}. Substantial Monte Carlo studies have also been performed for nonzero temperature bosonic multi-matrix theories, e.g. \cite{Azuma:2014cfa, Bergner:2019rca}.
However, recent work increasingly suggests that
the quantum structure of holographic quantum states --- revealed for instance in their entanglement \cite{Bianchi:2012ev, Faulkner:2013ana, Donnelly:2016auv,Harlow:2016vwg} --- plays a central role in the emergence of space. It therefore behooves us to find methods suitable for studying the zero temperature quantum states of multi-matrix quantum mechanics directly. Progress was made recently in this direction by using a neural network variational wavefunction \cite{Han:2019wue}. Here we describe a different approach.

Our work is directly inspired by a recent beautiful paper by Lin \cite{Lin:2020mme}, with a similar approach also being employed in \cite{ANDERSON2017702}. Lin's paper studied large $N$ matrix integrals, which is an easier problem than large $N$ quantum mechanics but shares important features. Positivity constraints and relations between correlation functions were shown to efficiently produce strong numerical bounds on correlation functions of matrix integrals. In the following we will show how this methodology can be adapted to the quantum mechanical problem.

{\it Bootstrapping the quantum anharmonic oscillator.---} We first illustrate the approach with
a warm-up example of a quantum anharmonic oscillator, with Hamiltonian
\begin{equation}
    H = p^2 + x^2 + g x^4 \,. \label{eq:hamil_osc}
\end{equation}
Here $[p, x] = -i$. Fig.~\ref{fig:osc} below shows the results for this case: strong constraints on the energy $E$ and expectation value $\langle x^2 \rangle$ of the ground state and first excited state. 

The first step is to relate the expectation values of different operators. We will obtain the recursion relation in (\ref{eq:osc_recur}) below. In energy eigenstates, for any operator $\ocal$, 
\begin{equation}
    \expect{[H, \ocal]} = 0. \label{eq:hcomm}
\end{equation}
For example, let $\ocal = x p$. Eq. (\ref{eq:hcomm}) is then the Virial theorem, $\expect{2 p^2} = \expect{2 x^2 + 4 g x^4}$. The energy is therefore
\begin{equation}
    E = 2 \expect{x^2} + 3 g \expect{x^4}. \label{eq:osc_e}
\end{equation}

More systematically, take $\ocal = x^s$ and $\ocal = x^t p$ in (\ref{eq:hcomm}) for integers $s, t \geq 0$. 
Commuting the operators $x, p$ with the identity $[p, x^r] = - i r x^{r- 1}$
and eliminating the terms with a single $p$ operator, we arrive at the relation
\begin{equation}
4 t \expect{x^{t - 1} p^2} = 8 g \expect{x^{t + 3}} + 4 \expect{x^{t + 1}} - t (t - 1) (t - 2) \expect{x^{t - 3}} \,. \label{eq:precur}
\end{equation}

In this single particle case is there is a strengthened version of (\ref{eq:hcomm}): $\expect{\ocal H} = E \expect{\ocal} \label{eq:heig}$. We emphasize (\ref{eq:hcomm}) instead because, as we will see later, it is more useful in the matrix case. Nonetheless, in the present anharmonic oscillator example, take $\ocal = x^{t - 1}$, so that
\begin{equation}
    \expect{x^{t - 1} p^2} = E \expect{x^{t - 1}} - \expect{x^{t + 1}} - g \expect{x^{t + 3}}. \label{eq:xp2}
\end{equation}
Plugging (\ref{eq:xp2}) into (\ref{eq:precur}) gives a recursive relation between expectation values of powers of $x$:
\begin{align}
    4 t E \expect{x^{t - 1}} &+ t (t - 1) (t - 2) \expect{x^{t - 3}} \nonumber \\
    &- 4 (t + 1) \expect{x^{t + 1}} - 4 g (t + 2) \expect{x^{t + 3}} = 0, \label{eq:osc_recur}
\end{align}
where $E$ is given by (\ref{eq:osc_e}). Also we know that $\expect{x^0} = 1$ and $\expect{x^t} = 0$ if $t$ is odd, so all expectation values of $x^t$ can be computed from $E$ and $\expect{x^2}$ with (\ref{eq:osc_recur}). 

With the recursion relation (\ref{eq:osc_recur}) at hand we move onto the second step. We wish to solve for $E$ and $\expect{x^2}$, the only two unknown variables, by bootstrapping.
\begin{figure}[t]
     \centering
    \includegraphics[width=0.45\textwidth]{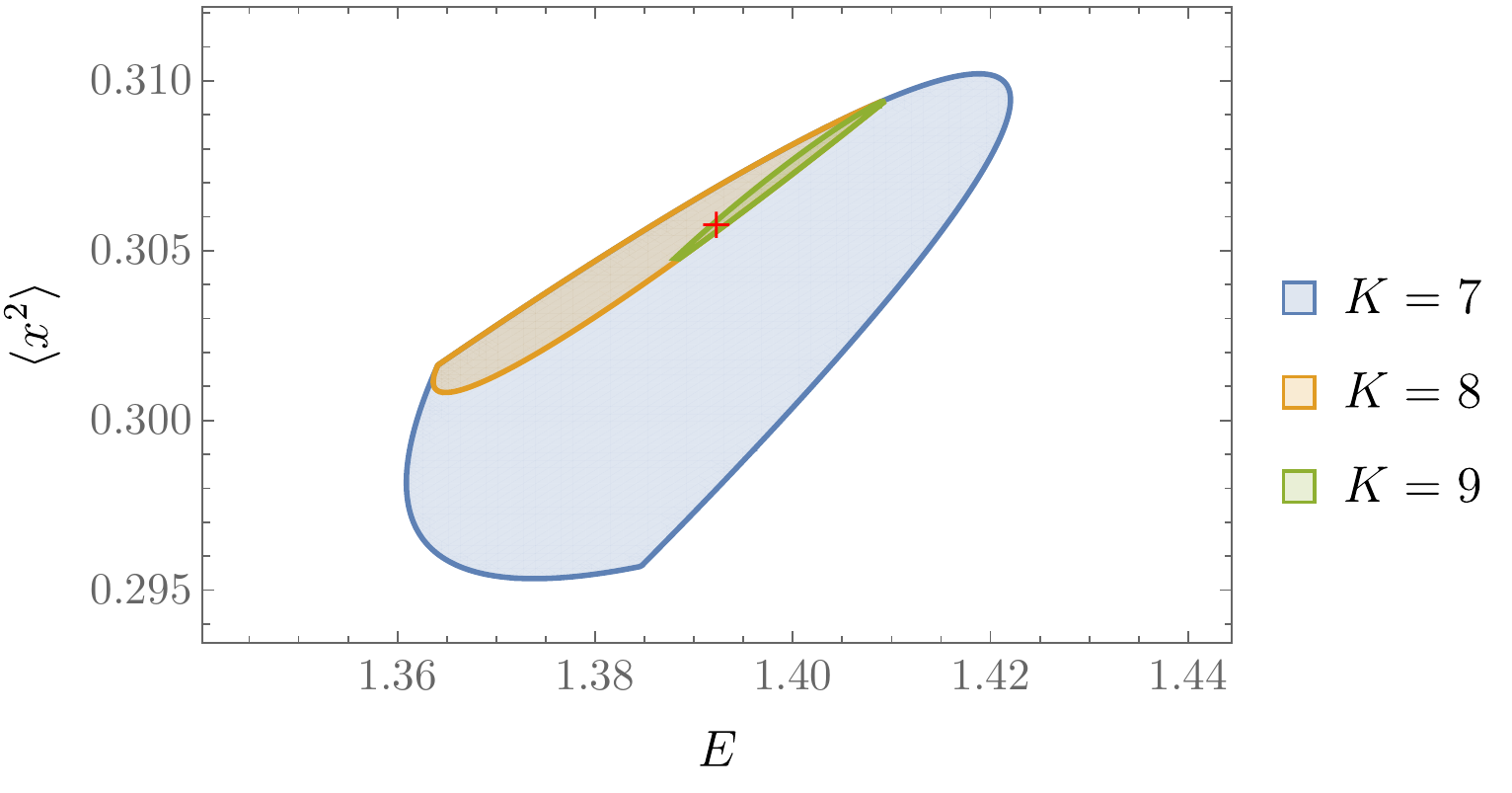}
    \includegraphics[width=0.45\textwidth]{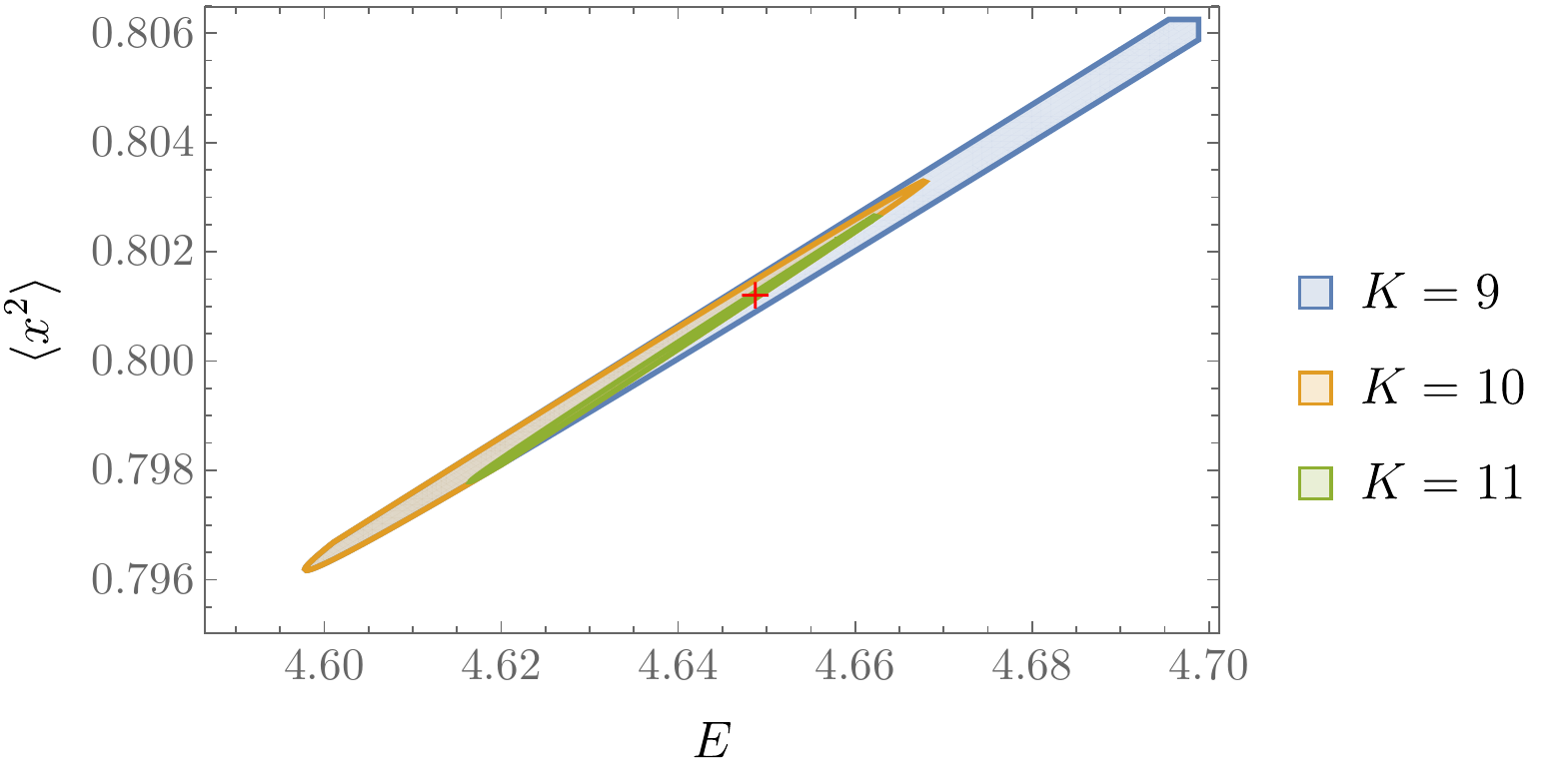}
    \caption{Bootstrap allowed region (shaded) for the anharmonic oscillator (\ref{eq:hamil_osc}) with $g = 1$. Upper plot: the allowed region for $(E, \expect{x^2})$ near the ground state solution (marked by the red cross) for different sizes of the bootstrap matrix $K = 7, 8, 9$; lower plot: the allowed region near the first excited state. }
    \label{fig:osc}
\end{figure}
This step works as in \cite{Lin:2020mme}.
The basic positivity constraint is that
\begin{equation}
    \expect{\ocal^\dagger \ocal} \geq 0\,, \qquad \forall \ocal = \sum_{i = 0}^K c_i x^i\,, \label{eq:positivity}
\end{equation}
which means that the matrix $\mcal$ of size $(K + 1) \times (K + 1)$,
$\mcal_{i j} = \expect{x^{i + j}}$, should be positive semidefinite. The constraint becomes stronger as we increase $K$, thus enlarging the space of trial operators. For a given $K$ and test values of $E$ and $\langle x^2 \rangle$, the $\mcal_{i j}$ can be computed using the recursion relation (\ref{eq:osc_recur}). The bootstrap consists in scanning over these test values, computing the eigenvalues of the matrix $\mcal$, and thereby determining if positivity excludes the test values as inconsistent.

The result is shown in Fig.\,\ref{fig:osc}. Even for moderate $K$ the values of $E$ and $\expect{x^2}$ are determined quite accurately. The region of allowed values splits into a discrete set of islands. These converge to the spectrum of the Hamiltonian in the limit $K \to \infty$ \footnote{If $\langle I \rangle = 1$, $\langle \ocal^\dagger \rangle = \langle \ocal \rangle^*$ and $\langle \ocal^\dagger \ocal \rangle \geq 0$ for {\it all} operators $\ocal$, then $\langle \ocal \rangle = \tr (\rho \ocal)$ for some quantum state $\rho$. If furthermore $\expect{\ocal H} = E \expect{\ocal}$, then $\rho$ must be an eigenstate with energy $E$. Therefore as $K \to \infty$, wherein the constraints are indeed imposed for all operators, the allowed region of energies necessarily shrinks to the spectrum of the Hamiltonian, with $\langle \ocal \rangle$ the expectation value in energy eigenstates}. Higher energy states require more constraints to be computed accurately.

{\it One matrix quantum mechanics.---}
Now we generalize the bootstrap method to matrix quantum mechanics at $N = \infty$. 
The momentum operators can no longer be eliminated explicitly in favor of the energy, and we do not use a closed form recursion relation for all expectation values. However, the energy and expectation values of short operators can still be efficiently constrained.

\begin{figure}[t]
    \centering
    \includegraphics[width=0.45\textwidth]{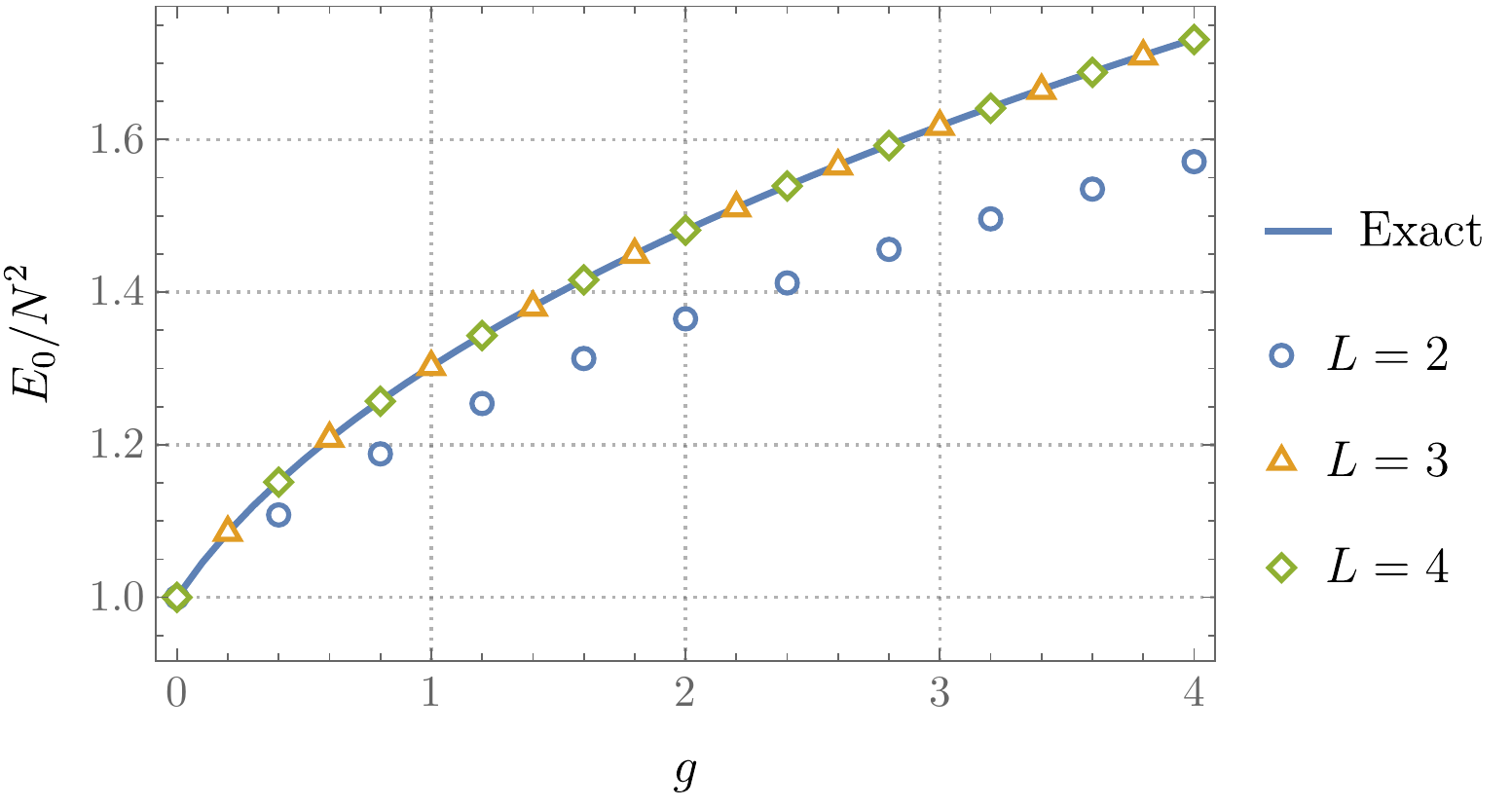}
    \includegraphics[width=0.45\textwidth]{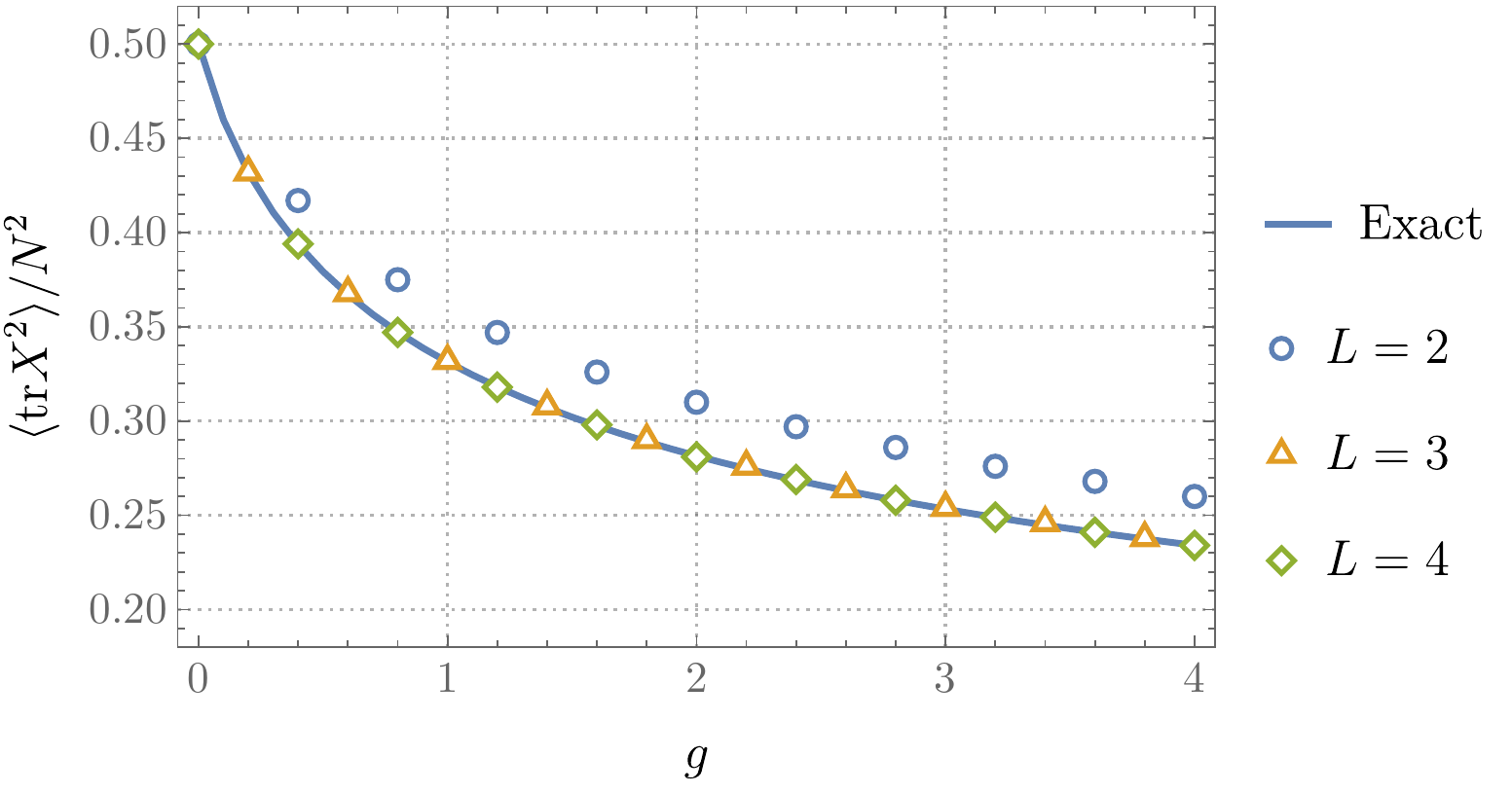}
    \caption{One matrix quantum mechanics bootstrap for the Hamiltonian (\ref{eq:h}). $L$ is the maximal length of trial operators. Upper: The markers show the minimal energies allowed by the bootstrap constraints, in comparison with the exact ground state solution. Lower: the expectation values of $\tr X^2$, for the minimal energy parameters found in the upper plot.}
    \label{fig:one-mat}
\end{figure}

Consider the single-matrix quantum mechanics with
\begin{equation}
    H = \tr P^2 + \tr X^2 + \frac{g}{N} \tr X^4, \label{eq:h}
\end{equation}
where $P$ and $X$ are $N$-by-$N$ Hermitian matrices with quantum commutators $[P_{i j}, X_{k l}] = - i \delta_{i l} \delta_{j k}$. The theory (\ref{eq:h}) can be solved by mapping onto $N$ free fermions \cite{Brezin:1977sv}. The bootstrap reproduces this solution in Fig. \ref{fig:one-mat}.

Operator expectation values are related by symmetries. In the following, denote $\expect{\ocal} = \tr \rho \,\ocal$. If the state $\rho$ commutes with the Hamiltonian then
\begin{equation}
    \expect{[H, \ocal]} = 0, \quad \forall \ocal. \label{eq:one_hcomm}
\end{equation}
For example, $\rho$ could be a pure energy eigenstate or a mixed thermal state. Choosing $\ocal = \tr X P$,
\begin{equation}
    2 \expect{\tr P^2} = 2 \expect{\tr X^2} + \frac{4 g}{N} \expect{\tr X^4}. \label{eq:one_virial}
\end{equation}

The $SU(N)$ symmetry of (\ref{eq:h}) has generators
\begin{equation}\label{eq:GGG}
    G = i [X, P] + N I \,.
\end{equation}
The final identity piece ensures that $\expect{\tr G} = 0$, with the operator ordering $[X,P] = XP - PX$ in (\ref{eq:GGG}). In gauged matrix quantum mechanics, physical states must be invariant under this symmetry. In particular,
\begin{equation}
   \expect{\tr G \ocal} = 0, \quad \forall \ocal_{i j}. \label{eq:one_gcons}
\end{equation}
For example, $\expect{\tr G} = 0$ implies
$\expect{\tr X P} - \expect{\tr P X} = i N^2$. Combining this constraint with $\expect{[H, \tr X^2]} = 0$ gives
\begin{equation}
    \expect{\tr X P} = - \expect{\tr P X} = \frac{i N^2}{2}. \label{eq:one_xp}
\end{equation}

Cyclicity of the trace gives another set of relations between operators. Commuting quantum operators may be necessary in applying the cyclic formula. For example, using large $N$ factorization to leading order in $N \to \infty$,
\begin{equation}
    \expect{\tr X P^3} = \expect{\tr P^3 X} + 2 i N \expect{\tr P^2} + i \expect{\tr P} \expect{\tr P}. \label{eq:cycl}
\end{equation}
Equations (\ref{eq:one_hcomm}), (\ref{eq:one_gcons}), cyclicity of the trace, and reality conditions $\expect{\ocal^\dagger} = \expect{\ocal}^*$ generate all relations between expectation values that we will use for the bootstrap. 

As a mini-bootstrap example, consider trial operators $I, X, X^2$ and $P$. From the condition (\ref{eq:positivity}),
the following bootstrap matrix should be positive semidefinite:
\begin{equation}
    \begin{array}{c|cccc}
          & I      &     X^2 &      X &      P  \\ \hline
        I & \expect{\tr I}  & \expect{\tr X^2} &  0 &  0  \\
        X^2 & \expect{\tr X^2}  &\expect{\tr X^4} &0 & 0  \\
        X& 0 & 0& \expect{\tr X^2} & \expect{\tr XP} \\
        P & 0  &0 & \expect{\tr PX} & \expect{\tr P^2}
    \end{array}\label{eq:easy}
\end{equation}
Trial operators are built from both $X$ and $P$. 
The expectation value for an odd number of matrices vanishes. Positivity of (\ref{eq:easy}) implies
\begin{align}
    \expect{\tr X^2} \geq 0, \quad N \expect{\tr X^4} \geq \expect{\tr X^2}^2, \nonumber \\
    \expect{\tr X^2} \left(\expect{\tr X^2} + \frac{2 g}{N} \expect{\tr X^4}\right) \geq \frac{N^4}{4}, \label{ineq:one}
\end{align}
where equations (\ref{eq:one_virial}) and (\ref{eq:one_xp}) are used. The inequalities (\ref{ineq:one}) are the bootstrap constraints in this simple example. At $g = 0$, $\expect{\tr X^2} = \frac{1}{2} N^2$ and $ \expect{\tr X^4} = \frac{1}{2} N^3$, so the last inequality in (\ref{ineq:one}) is saturated and the other two are not.

The bootstrap constraints become stronger as we include more trial operators. Firstly, take all possible strings of $X$ and $P$ of length $\leq L$, and write down the matrix analogous to (\ref{eq:easy}). This matrix must be positive semidefinite. Secondly, regard each of the $\sim 2^{2L}$ entries in the matrix as a variable (which is the expectation value of a single-trace operator with length $\leq 2 L$), and write down the equalities between them following from (\ref{eq:one_hcomm}), (\ref{eq:one_gcons}),
cyclicity of the trace, $\expect{\ocal^\dagger} = \expect{\ocal}^*$ and that the expectation value of an odd number of matrices vanishes. The technical implementation of these constraints, as well as the minimization described in the following paragraph, is detailed in \footnote{\label{foot} See supplementary material below.}.

Unlike in the single-particle case, we do not necessarily require that the state be an energy eigenstate and the energy $E$ does not appear explicitly in the bootstrap constraints. At infinite $N$ the matrix quantum mechanics has a continuous spectrum and therefore 
we proceed to use gradient descent to minimize the energy in the allowed region of expectation values. In this way we obtain a lower bound on the ground state energy of the theory. The result is a lower bound because certainly the true ground state energy is allowed, and hence above the minimal allowed energy that we find. In Fig.\,\ref{fig:one-mat} we observe that the lower bound is very close to the true ground state value, already for $L=3$, and other observables, such as $\expect{\tr X^2}$, are also solved accurately. 

{\it Two matrix quantum mechanics.---}
One matrix quantum mechanics are tractable analytically as
one can diagonalize the matrix. This is not the case for
multi-matrix quantum mechanics.
In the following we illustrate how bootstrap methods can successfully be used for such theories, focussing on a relatively simple two-matrix quantum mechanics with a global $O(2)$ symmetry (in addition to the large $N$ gauge symmetry). The Hamiltonian is
\be
H = \tr \left( P_X^2 + P_Y^2 + m^2(X^2 + Y^2) - g^2 [X,Y]^2 \right), \label{eq:h_two}
\ee
with $X$ and $Y$ being $N$-by-$N$ Hermitian matrices, with conjugate momenta $P_X$ and $P_Y$, and $m$ and $g$ coupling constants. This theory is not exactly solvable. An early discussion of the massless ($m=0$) limit of the theory is \cite{hoppe1982quantum}. By rescaling the matrices we see that dimensionless physical quantities can only depend on the ratio $m^2/ g^{4/3}$.

\begin{figure}
    \centering
    \includegraphics[width=0.45\textwidth]{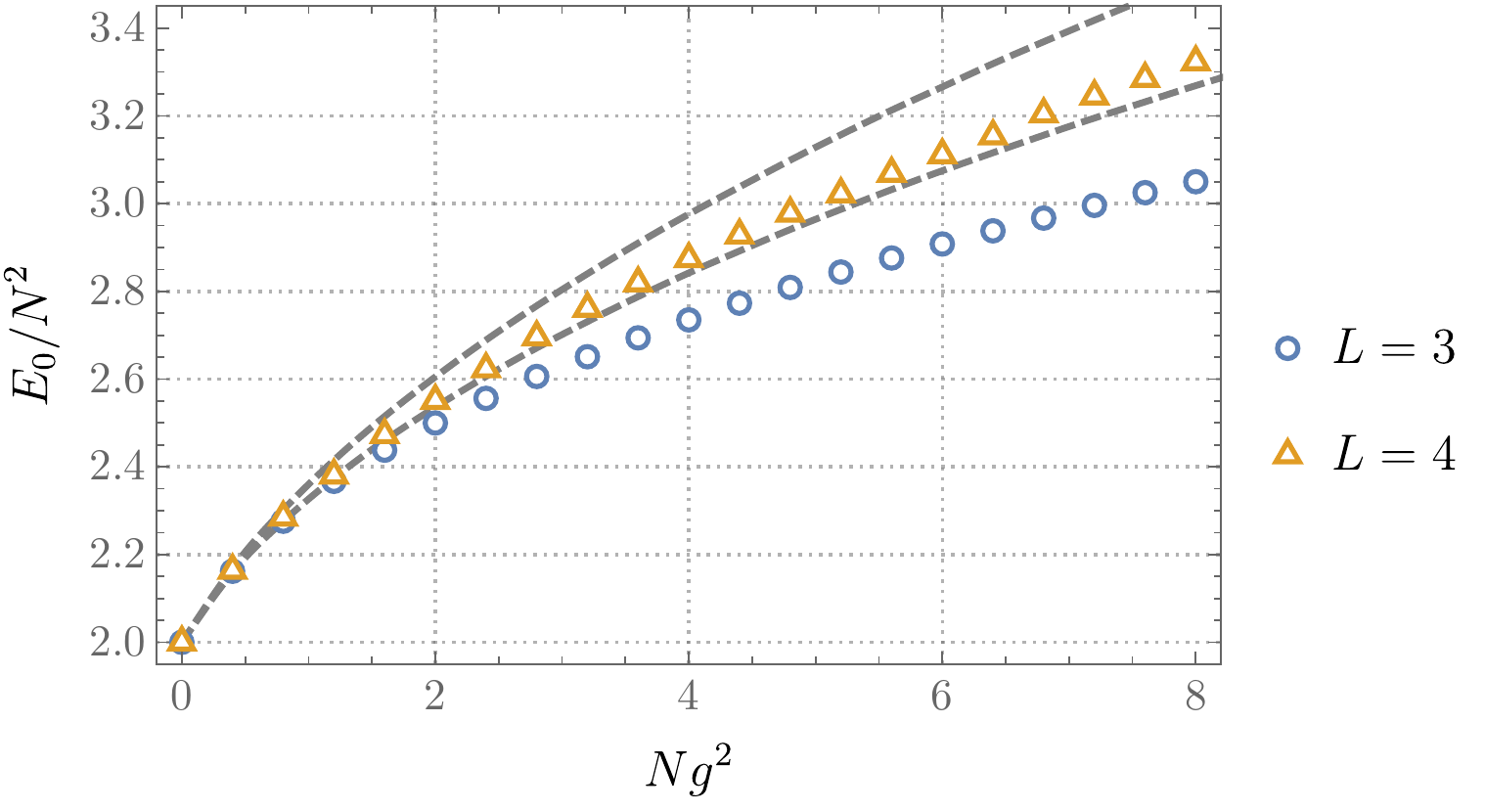}
    \includegraphics[width=0.45\textwidth]{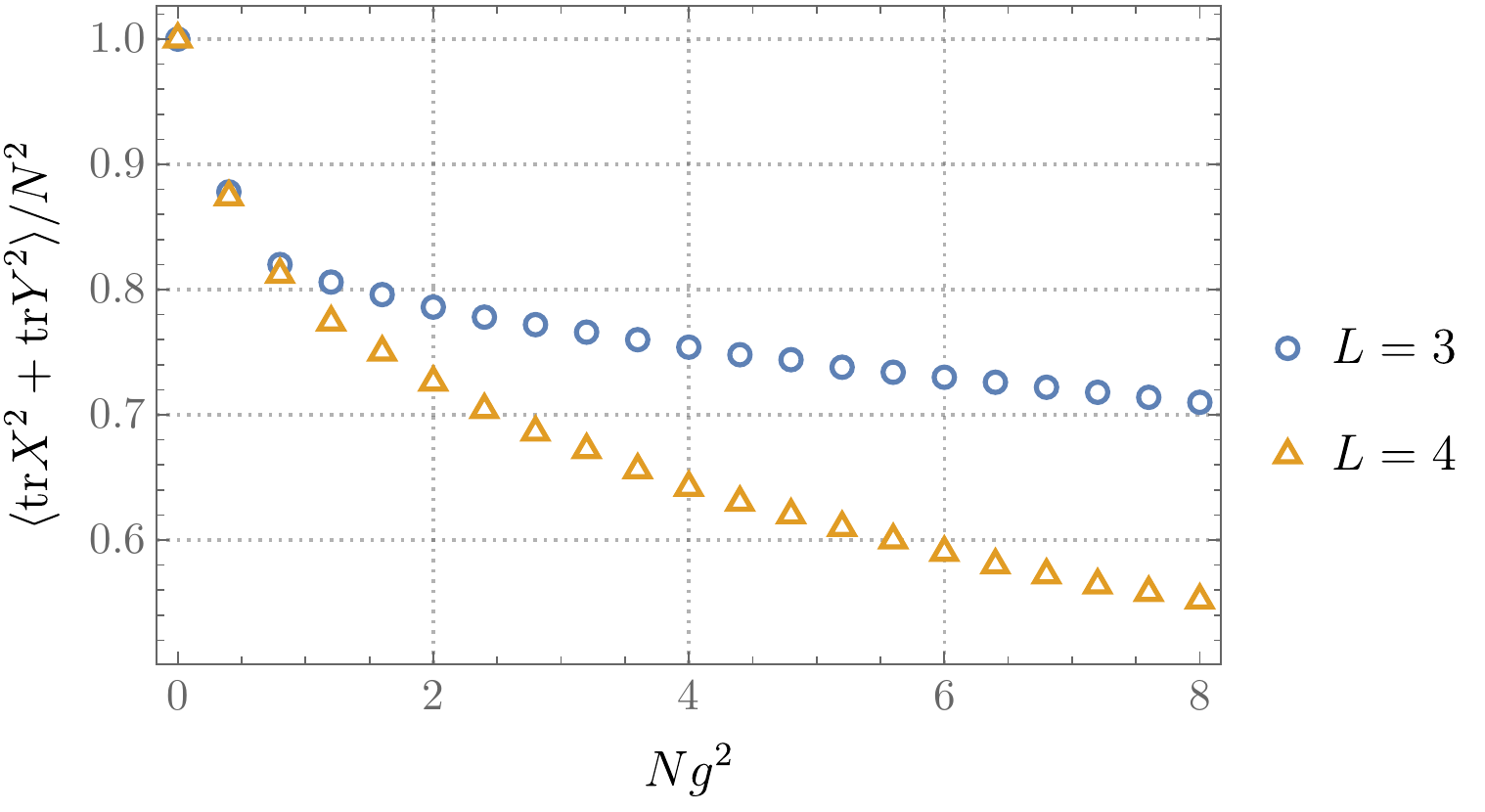}
    \includegraphics[width=0.45\textwidth]{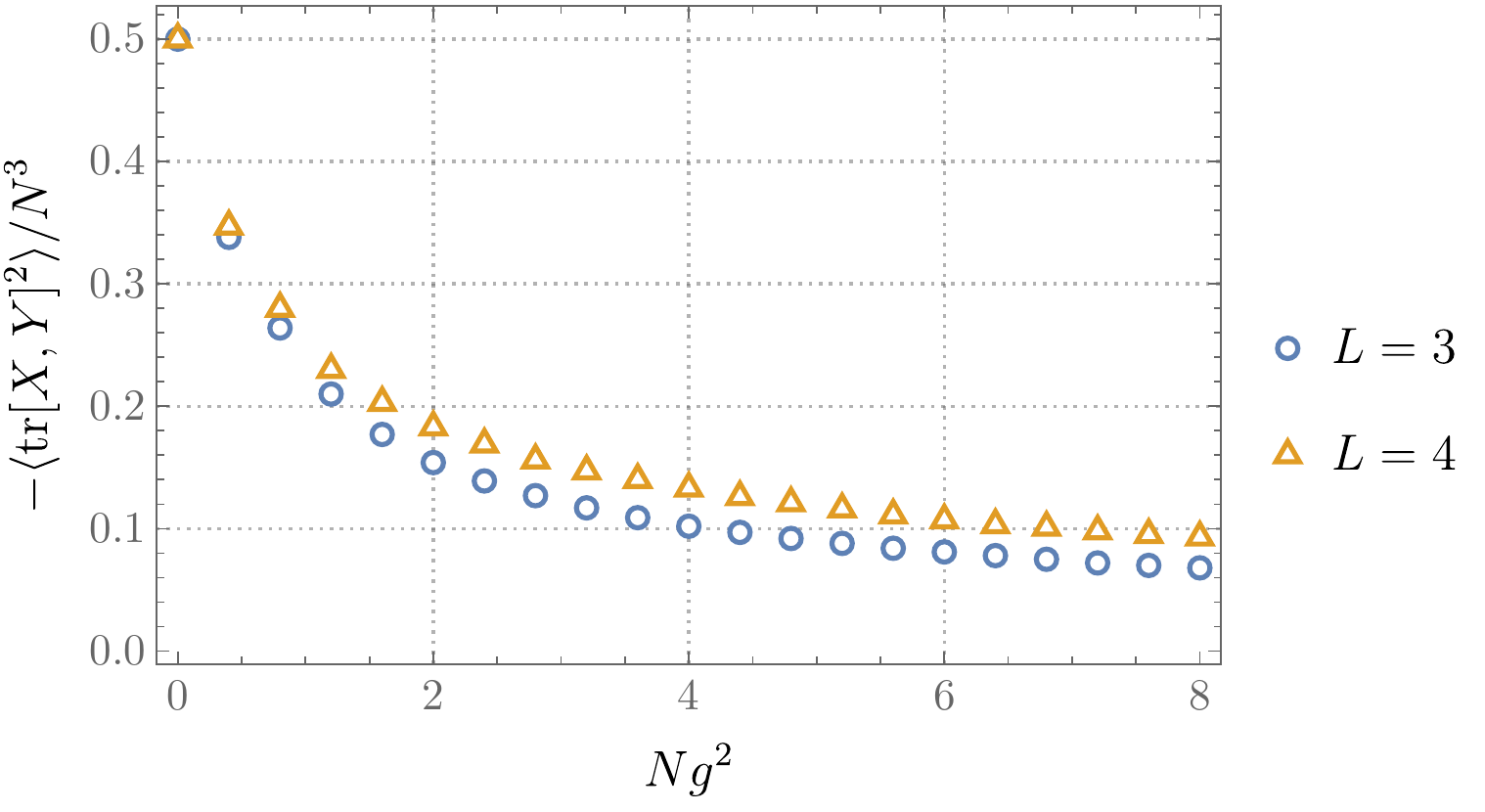}
    \caption{Minimal energy configuration in the bootstrap allowed region for $L = 3, 4$. The gray dashed curves are rigorous lower and upper bounds of the ground state energy from the Born-Oppenheimer approximation. In the plots we have set $m = 1$.}
    \label{fig:e_two_matrix}
\end{figure}

Imposing rotational invariance gives more relations between observables. We expect the ground state to be rotationally invariant. Rotations are generated by
\begin{equation}
    S = \tr (X P_Y - Y P_X) \,.
\end{equation}
For states $\rho$ with $[S,\rho] = 0$, including eigenstates of $S$,
\begin{equation}
    \expect{[S, \ocal]} = 0, \quad \forall \ocal. \label{eq:rot}
\end{equation}
Thus in the two matrix quantum mechanics, equations (\ref{eq:one_hcomm}), (\ref{eq:one_gcons}), (\ref{eq:rot}), cyclicity of the trace, and $\expect{\ocal^\dagger} = \expect{\ocal}^*$ will be used to generate all equations between expectation values that we will use. The bootstrap then proceeds in exactly the same way as for the case of a single matrix, now with $\sim 4^{2L}$ variables prior to imposing constraints. 
The results for the ground state energy, $\langle \tr X^2 + \tr Y^2 \rangle$ and $\langle \tr [X,Y]^2 \rangle$ are in Fig. \ref{fig:e_two_matrix}. The Virial theorem relates these: $E_0 = 2 m^2 \langle \tr X^2 + \tr Y^2 \rangle - 3 g^2 \langle \tr [X,Y]^2 \rangle$.

In order to corroborate the accuracy of the $L=4$ results, we obtain rigorous upper and lower bounds on the true ground state energy using a  Born-Oppenheimer wavefunction. We see in Fig. \ref{fig:e_two_matrix} that the $L=4$ bootstrap results indeed lie within a narrow window allowed by these bounds. We briefly describe the wavefunction in the following paragraph, with details given in \cite{Note2}. As further evidence that the $L=4$ bootstrap results are close to convergence, we compare our results to existing low temperature Monte Carlo simulations of the massless theory. At large $g$, $E_0/N^2 \approx 1.40 \, (N g^2)^{1/3} + 1.01\, m^2 / (N g^2)^{1/3}$ from data in Fig. \ref{fig:e_two_matrix}. The factor of 1.40 agrees precisely with the Monte Carlo result in \cite{Morita:2020liy}, corresponding to the value of $0.70$ in the conventions of that paper. An analogous fit gives the leading order behavior $\langle \tr X^2 + \tr Y^2 \rangle/N^2 \approx 1.22/(N g^2)^{1/3}$. The numerical factor here is close to the Monte Carlo result of $1.15$ in \cite{Morita:2020liy}.

The $SU(N)$ gauge invariance allows us to diagonalize one of the two matrices, say $X$. Let the eigenvalues be $x_i$. The Hamiltonian for the entries $y_{ij}$ of the remaining matrix is a sum of harmonic oscillators, with frequencies $\omega_{ij}^2 = m^2 + g^2(x_i - x_j)^2$. We can therefore write down a Born-Oppenheimer wavefunction in which these oscillators are placed in their ground state:
\be\label{eq:BO}
\Psi(X,Y) = \psi(x_i) \prod_{i,j = 1}^N (2 \w_{ij} / \pi)^{1/4} e^{-\frac{1}{2}\w_{ij}|y_{ij}|^2} \,.
\ee
That is, the $y_{ij}$ are treated as `fast' compared to the eigenvalues $x_i$. Born-Oppenheimer wavefunctions lead to both upper and lower bounds on the ground state energy. The upper bound follows from treating the wavefunction as a variational ansatz. The lower bound is obtained by finding the ground state of the eigenvalues in an effective potential due to the zero point energy of the $y_{ij}$ oscillators.
The advantage of the form (\ref{eq:BO}) is that computing the upper and lower bounds reduces to a solvable single-matrix large $N$ eigenvalue problem. In Fig. \ref{fig:e_two_matrix} we see that the bounds following from the wavefunction (\ref{eq:BO}) turn out to be remarkably tight.

From the results in Fig.\,\ref{fig:e_two_matrix} one can verify that the ratio $N \tr [X,Y]^2/(\tr X^2)^2$ tends to a nonzero constant at large $N g^2$. This means that the matrices do not commute in this limit. This can be constrasted with the analogous two matrix integral, with no time, that does become commuting at large $N g^2$ \cite{Berenstein:2008eg}. This is consistent with the fact that the two matrix integral diverges in the massless limit \cite{Krauth:1998xh, Krauth:1998yu}, as the eigenvalues spread far apart along the classically flat directions of the potential due to commuting matrices, while the massless matrix quantum mechanics still has a discrete spectrum of normalizable states \cite{SIMON1983209}. 

{\it Final comments.---}
In summary, we have introduced a systematic numerical method to obtain energies and expectation values of large $N$ matrix quantum mechanics states.
The method involves establishing relationships between expectation values and then imposing positivity of a certain matrix of expectation values, in the spirit of \cite{Lin:2020mme}. In Fig. \ref{fig:one-mat} we see that the known analytic results for one-matrix large $N$ quantum mechanics are readily reproduced. In Fig. \ref{fig:e_two_matrix} we have obtained new results for the ground state energy and expectation values of a two-matrix large $N$ quantum mechanics.

The extension to more matrices should be possible with increased computing power or perhaps by optimizing the algorithm. Looking at supersymmetric states in supersymmetric theories may allow for stronger relationships between expectation values, using the supersymmetry generators. Both more matrices and supersymmetry will of course be necessary to tackle the full blown BFSS and BMN theories. Finally, extensions to Gibbs states (or, to high energy eigenstates) may allow nonzero temperature quantum physics to be accessed with our bootstrap methods. 
This could give an alternative probe of the thermal phase transitions studied via Monte Carlo in e.g. \cite{Azuma:2014cfa, Bergner:2019rca}, as well as a new window onto black hole microstates.

\section*{Acknowledgements}

This work arose from discussions with Edward Mazenc and Daniel Ranard, who also collaborated on the early stages of the project. JK is supported by the Simons Foundation. SAH is partially supported by DOE award de-sc0018134 and by a Simons Investigator award.

\newpage

\onecolumngrid

\appendix

\setcounter{equation}{0}

\renewcommand{\theequation}{S\arabic{equation}}
\renewcommand\thefigure{S\arabic{figure}}

{\begin{center} {\bf \large Supplementary Material for `Bootstrapping Matrix Quantum Mechanics'} \\
\vspace{0.3cm}
Xizhi Han, Sean A. Hartnoll, Jorrit Kruthoff \end{center}}

\bigskip

\section{Born-Oppenheimer wavefunction}
\label{app:BO}

This section gives details of computations involving a Born-Oppenheimer wavefunction for the two matrix quantum mechanics:
\begin{equation} \label{eq:h_twoB}
    H = \tr \left( P_X^2 + P_Y^2 + m^2(X^2 + Y^2) - g^2 [X,Y]^2 \right).
\end{equation}
The role of this wavefunction is to give a lower and an upper bound on the actual ground state energy. This gives a check on the accuracy of our numerical bootstrap in this case.
The results of this section are the effective Hamiltonians (\ref{Hred}) and (\ref{HBOred}) for the eigenvalues of one of the two matrices. These will be solved in the following section \ref{app:collective}, giving the upper and lower bounds respectively.

The wavefunction that we are searching for is a complex function $\Psi(X, Y)$ of Hermitian matrices $X$ and $Y$. The state should be $SU(N)$ gauge invariant and hence for any unitary matrix $W \in SU(N)$,
\begin{equation} \label{eq:gauge_inv}
    \Psi(X, Y) = \Psi(W X W^{-1}, W Y W^{-1}). 
\end{equation}
It will be convenient to parametrize such a state with the following set of variables: a diagonal real matrix $x_i$, a Hermitian matrix $y_{ij}$ and a unitary matrix $U \in SU(N)$, such that
\begin{equation}\label{eq:change}
    X = U \mathrm{diag}(x_i) U^{-1}, \quad Y = U y U^{-1} \,.
\end{equation}
In these variables we can write down the following
Born-Oppenheimer ansatz, in which the $y_{ij}$ oscillators are put in their ground state for a fixed configuration of eigenvalues $x_i$:
\be\label{eq:BOapp}
\Psi(X,Y) = \psi(x_i) \phi(x_i, y_{ij}), \quad \phi(x_i, y_{i j}) = \prod_{i,j = 1}^N (2 \w_{ij} / \pi)^{1/4} e^{-\frac{1}{2}\w_{ij}|y_{ij}|^2} \,,
\ee
with $\w_{ij}^2 = m^2 + g^2(x_i - x_j)^2$. Equation (\ref{eq:BOapp}) defines a gauge invariant wavefunction by specifying its values on the gauge slice where $X$ is diagonal. However, we should check that (\ref{eq:BOapp}) is well-defined because (\ref{eq:change}) does not uniquely determine $x_i$ and $y_{i j}$ as a function of $X$ and $Y$. Indeed, there is a residual $U(1)^{N - 1}$ gauge symmetry after fixing $X$ to be diagonal: if we choose $U = \mathrm{diag}(\exp i \theta_i)$ in (\ref{eq:change}), $X = \mathrm{diag}(x_i)$ but $Y_{i j} = y_{i j} \exp i (\theta_i - \theta_j)$. Because (\ref{eq:BOapp}) is invariant under this residual gauge symmetry as well, $\Psi(X, Y)$ in (\ref{eq:BOapp}) is well-defined.  

To obtain a variational upper bound, we wish to find an effective Hamiltonian for the `slow' $x_i$ degrees of freedom that calculates the expectation value of the full Hamiltonian (\ref{eq:h_twoB}) in the variational state (\ref{eq:BOapp}). The expectation value of the Hamiltonian in the state $\Psi$ consists of a kinetic part and a potential part:
\begin{equation}
    \langle \Psi | H |\Psi \rangle = \int d X d Y\, \Psi^*(X, Y) (H_{\rm kin} + H_{\rm pot}) \Psi(X, Y) \,.
\end{equation}
We discuss these in turn. The kinetic energy is
\begin{equation} \label{eq:kin_e}
    \langle \Psi | H_{\rm kin} |\Psi \rangle = \sum_{i, j = 1}^N \int d X d Y\, \left(\left|\frac{\partial \Psi(X, Y)}{\partial X_{i j}}\right|^2 + \left|\frac{\partial \Psi(X, Y)}{\partial Y_{i j}}\right|^2\right).
\end{equation}
Here $\partial / \partial X_{i j} = \frac{1}{2}(\partial / \partial \mathrm{Re} X_{i j}  - i\partial / \partial \mathrm{Im} X_{i j})$ are complex derivatives because the matrices are Hermitian. Because the kinetic energy operator is also gauge invariant, the integrand in (\ref{eq:kin_e}) is constant along gauge orbits. So it suffices to evaluate it on the gauge slice where $U$ in (\ref{eq:change}) is the identity. Then by the chain rule and (\ref{eq:change}), at $U = I$,
\begin{equation} \label{eq:der_xy}
    \frac{\partial \Psi}{\partial x_i}= \frac{\partial \Psi}{\partial X_{i i}}, \quad \frac{\partial \Psi}{\partial y_{i j}} = \frac{\partial \Psi}{\partial Y_{i j}},
\end{equation}
and
\begin{equation}
    \frac{\partial \Psi}{\partial U_{i j}} =  (x_j - x_i) \frac{\partial \Psi}{\partial X_{i j}} + \sum_{m, n = 1}^N (\delta_{i m} y_{j n} - \delta_{j n} y_{m i}) \frac{\partial \Psi}{\partial Y_{m n}}.
\end{equation}
Because $\Psi$ is gauge invariant as in (\ref{eq:gauge_inv}), $\partial \Psi / \partial U = 0$ so for $i \neq j$,
\begin{equation}\label{eq:kin_sol}
    \frac{\partial \Psi}{\partial X_{i j}} = \frac{1}{x_i - x_j} \sum_{m, n = 1}^N (\delta_{i m} y_{j n} - \delta_{j n} y_{m i}) \frac{\partial \Psi}{\partial y_{m n}}.
\end{equation}
Plug (\ref{eq:der_xy}) and (\ref{eq:kin_sol}) into (\ref{eq:kin_e}) and evaluate the $y_{i j}$ integrals in the state (\ref{eq:BOapp}), 
\begin{align}
    \langle \Psi | H_{\rm kin} |\Psi \rangle = \int \Delta(x_i) d x_i \, \left( \sum_{i = 1}^N \left|\frac{\partial \psi}{\partial x_i}\right|^2 + |\psi|^2 \sum_{i, j = 1}^N \frac{\w_{i j}}{2} + |\psi|^2 \sum_{i,j,k = 1}^N \frac{(\omega_{i k} - \omega_{j k})^2}{4 \omega_{i k} \omega_{j k} (x_i - x_j)^2}  \right),
\end{align}
where $\D = \prod_{i<j} (x_i - x_j)^2$ is the usual Vandermonde determinant, with $d X d Y = \D d x_i d y_{i j}$. 

The potential term on the gauge slice $U = I$ is 
\begin{equation} \label{eq:hpot}
    H_{\rm pot} = \sum_{i = 1}^N m^2 x_i^2 + \sum_{i, j = 1}^N \w_{i j}^2 |y_{ij}|^2,
\end{equation}
and thus
\begin{equation}
    \langle \Psi | H_{\rm pot} | \Psi \rangle = \int \Delta(x_i) d x_i \, \psi^*(x_i) \left(\sum_{i = 1}^N m^2 x_i^2 + \sum_{i, j = 1}^N \frac{\w_{i j}}{2}\right) \psi(x_i).
\end{equation}
Overall the effective variational Hamiltonian on $x_i$, such that $\langle \Psi|H|\Psi \rangle = \langle\psi|H_\text{var}|\psi \rangle$, is therefore
\begin{align}\label{Hred}
H_{\rm var} = \sum_{i = 1}^N \left( -\frac{1}{\D}\frac{\partial}{\partial x_i}\left( \D \frac{\partial}{\partial x_i} \right) + m^2 x_i^2\right) + \sum_{i, j = 1}^N  \w_{ij} + \sum_{i,j,k = 1}^N \frac{(\omega_{i k} - \omega_{j k})^2}{4 \omega_{i k} \omega_{j k}(x_i - x_j)^2}  \,.
\end{align}

The choice of gauge and the form of the ansatz (\ref{eq:BOapp}) break rotational symmetry. We have done this because it has allowed the problem to be reduced to a single-matrix eigenvalue Hamiltonian (\ref{Hred}), which we will be able to solve explicitly. It is possible to restore rotational symmetry by acting on the wavefunction with the generator of rotations. This will not change the energy of the variational state.

From the variational principle we know that the ground state energy of the reduced Hamiltonian (\ref{Hred}) is an upper bound on the ground state energy of the original Hamiltonian (\ref{eq:h_twoB}). However, it is well-known that Born-Oppenheimer wavefunctions also give a lower bound on the ground state energy. In the present context (as we prove below) this means that if we drop the final term in (\ref{Hred}), the ground state energy of the Born-Oppenheimer Hamiltonian 
\be\label{HBOred}
H_{\rm BO} = \sum_{i = 1}^N \left( -\frac{1}{\D}\frac{\partial}{\partial x_i}\left( \D \frac{\partial}{\partial x_i} \right) + m^2 x_i^2\right) + \sum_{i, j = 1}^N \w_{ij} \,,
\ee
is a lower bound on the ground state energy of (\ref{eq:h_twoB}). 

A short proof of this fact is as follows: split the kinetic term into three parts $H_{\rm kin} = H_{\rm kin}^1 + H_{\rm kin}^2 + H_{\rm kin}^3$, where $H_{\rm kin}^1$ is the $\partial \Psi / \partial X_{ij}$ contribution in (\ref{eq:kin_e}), but where the derivative does not act on the $\phi$ part of the wavefunction (\ref{eq:BOapp}), $H_{\rm kin}^2$ is the $\partial \Psi / \partial X_{ij}$ contribution in (\ref{eq:kin_e}) minus  $H_{\rm kin}^1$, and $H_{\rm kin}^3$ is the remaining $\partial \Psi / \partial Y_{ij}$ term. Also split the potential term (\ref{eq:hpot}) into two pieces: $H_{\rm pot} = H_{\rm pot}^1 + H_{\rm pot}^2$, where $ H_{\rm pot}^1$ is the first sum in (\ref{eq:hpot}) and  $H_{\rm pot}^2$ the second. Now note that $\phi(x_i, y_{i j})$ in (\ref{eq:BOapp}) is the ground state of the harmonic oscillator Hamiltonian $H_{\rm kin}^3 + H_{\rm pot}^2$ and that $H_{\rm kin}^2$ is positive semidefinite, so for any gauge invariant state $\Phi(x_i, y_{i j})$,
\begin{align}
    \langle \Phi | H | \Phi \rangle &\geq \langle \Phi | H_{\rm kin}^1 + H_{\rm kin}^2 + H_{\rm pot}^1 + E_{\rm BO}(x_i) | \Phi\rangle \nonumber \\
    &\geq \langle \Phi | H_{\rm kin}^1 + H_{\rm pot}^1 + E_{\rm BO}(x_i) | \Phi\rangle = \langle \Phi | H_{\rm BO} | \Phi\rangle,
\end{align}
where $E_{\rm BO}(x_i) = \sum_{i, j = 1}^N \w_{ij}$ is the ground state energy of the harmonic oscillator Hamiltonian for $y_{ij}$'s:
\begin{equation}
    H_{\rm kin}^3 + H_{\rm pot}^2 = - \sum_{i, j = 1}^N \frac{\partial^2}{\partial y_{i j} \partial y_{j i}} + \sum_{i, j = 1}^N \w_{i j}^2 |y_{ij}|^2.
\end{equation}

\section{Large $N$ collective field solution}
\label{app:collective}

In this section we solve for the ground state energies of the effective eigenvalue Hamiltonians (\ref{Hred}) and (\ref{HBOred}), using the large $N$ collective field method. We thereby obtain an upper and a lower bound for the ground state energy of (\ref{eq:h_twoB}). 
As is well known, at large $N$ the collective field of eigenvalues
\be
\rho(x) = \sum_{i=1}^N \delta(x - x_i) \,,
\ee
becomes classical. We can follow the established steps \cite{Das:1990kaa} to obtain the energy as a functional of this collective field. To obtain the Hamiltonian for $\rho(x)$ we must relate the derivative $\pa_{x_i}$ to the conjugate collective variable $\pi(x) = -i\delta/\delta \rho(x)$. The chain rule shows that
\be
\partial_{x_i} = i \pi'(x_i)\,, \qquad \partial_{x_i}^2 = i \pi''(x_i) - \pi'(x_i)^2 \,.
\ee
Plugging these into \eqref{HBOred} and defining 
\be
\rho_H(x) = \mathcal{P} \int dy \frac{\rho(y)}{x-y},
\ee
where $\mathcal{P}$ denotes taking the principal value, one finds 
\be\label{HredColl}
H_{\rm BO} = \int dx \rho(x) \left[ \pi'(x)^2 - 2i \rho_H(x) \pi'(x)  + V(x) \right] \,,
\ee
with
\be
V(x) = m^2 x^2 + \int dy \rho(y) \sqrt{m^2 + g^2(x - y)^2}  \,. \label{eq:vv}
\ee
We also used the fact that
\be
\mathcal{P} \int dx dy \rho(x)\rho(y) \frac{\pi'(x)}{x-y} = \sum_{i\neq j} \frac{\pi'(x_i)}{x_i - x_j} + \frac{1}{2}\int dx \rho(x) \pi''(x) \,.
\ee

The Hamiltonian in \eqref{HredColl} is not manifestly Hermitian. This can be cured by performing a canonical transformation that shifts $\pi'$ by $i \rho_H$, resulting in the Hamiltonian,
\be
H_{\rm BO} = \int dx \rho(x) \left[ \pi'(x)^2 + \rho_H(x)^2 + V(x) \right].
\ee
With this Hamiltonian we can straightforwardly compute the ground state energy and certain observables in the ground state. At large $N$ the eigenvalue distribution becomes classical and hence the momentum $\pi(x)$ vanishes in the ground state. Therefore it is sufficient to minimize the potential energy functional. Using the identity
\be
\int dx \rho(x)\rho_H(x)^2 = \frac{\pi^2}{3} \int dx \rho(x)^3 \,,
\ee
(here $\pi$ is the irrational number, not the conjugate momentum) this can be written as
\be
E_{\rm BO}[\rho] = \int dx \rho(x) \left( \frac{\pi^2}{3} \rho(x)^2 + m^2 x^2 \right) + \int dx dy \rho(x) \rho(y) \omega(x,y)  \,, \label{eq:Enap}
\ee
with
\be
\omega(x,y) = \sqrt{m^2 + g^2 (x-y)^2} \,.
\ee
Equation (\ref{eq:Enap}) must be minimized subject to the normalization constraint 
$\int dx \rho(x) = N$ and the constraint that $\rho(x)$ be pointwise non-negative. In the large $N$ limit, this normalization combined with balancing the terms in the energy functional and taking the mass to be fixed at order one (recall that the mass can be removed by rescaling the matrices) requires the scaling
\be
x \sim N^{1/2} \,, \qquad \rho \sim N^{1/2} \,, \qquad g^2 \sim \frac{1}{N} \,.
\ee
This is the familiar large $N$ scaling of these quantities. In particular the 't Hooft coupling $\lambda = g^2 N$ is finite in this limit.

The minimization of (\ref{eq:Enap}) is straightforward to perform numerically, by discretizing the integral. With the numerical solution at hand one can evaluate the energy $E_{\rm BO}$ of the state. These results are shown in Fig. \ref{fig:e_two_matrix} in the main text.

Similarly we can minimize the effective variational Hamiltonian (\ref{Hred}) to obtain an upper bound on $E_0$.
The steps are the same as above, and the functional to minimize is now
\begin{align}
E_{\rm var}[\rho] & = \int dx \rho(x) \left( \frac{\pi^2}{3} \rho(x)^2 + m^2 x^2 \right) + \int dx dy \rho(x) \rho(y) \omega(x,y)  \nonumber \\
& + \int d x dy d z \rho(x) \rho(y) \rho(z) \frac{\left(\omega(x,z) - \omega(y,z)\right)^2}{4 \omega(x,z) \omega(y,z) (x - y)^2} \,. \label{eq:Enap2}
\end{align}

As discussed in section \ref{app:BO}, we expect that the true ground state energy $E_0$ is bounded above and below as
\be\label{eq:ine}
E_0^\text{low} \equiv \min_\rho E_\text{BO}[\rho] \; \leq \; E_0 \; \leq \; \min_\rho E_\text{var}[\rho] \equiv E_0^\text{high} \,.
\ee
We can verify explicitly that these inequalities are obeyed in perturbation theory in small $g^2 N$. The ground state energy of the full Hamiltonian \eqref{eq:h_twoB} may be evaluated using standard quantum mechanical perturbation theory directly. The functionals $E_\text{BO}[\rho]$ and $E_\text{var}[\rho]$ are minimized within perturbation theory by a distribution of the form $\rho(x) = \sqrt{x_\star^2 - x^2} P(x)$, with $P(x)$ a polynomial (whose degree increases order by order in perturbation theory). At large $N$ we obtain
(with $\lambda = N g^2$ and $m=1$)
\begin{align}
\frac{E_0^{\rm low}}{N^2} & = 2 + \frac{1}{2} \lambda - \frac{7}{16}\lambda^2 + \frac{59}{64}\lambda^3 + \cdots \,, \\
\frac{E_0}{N^2} & = 2 + \frac{1}{2} \lambda - \frac{11}{32} \lambda^2 + \frac{137}{256} \lambda^3 + \cdots, \\
\frac{E_0^{\rm high}}{N^2} & = 2 + \frac{1}{2} \lambda - \frac{1}{4} \lambda^2 + \frac{3}{64} \lambda^3 + \cdots \,. 
\end{align}
In these expressions we see that the Born-Oppenheimer results only start to differ from the full answer at order $\lambda^2$ and that the inequalities (\ref{eq:ine}) are obeyed. Similar perturbative expansions have previously been considered at nonzero temperature in \cite{Aharony:2005ew}. The opposite limit of $\lambda \to \infty$ should approach the massless ($m=0$) result. It is simple to evaluate the lower bound in this limit. With $m=0$ and $\l = 1$ we find $E_0^{\rm low}/N^2 \approx 1.308$. This is indeed lower than the Monte Carlo result of $E_0^{\rm MC}/N^2 \approx 1.40$ for the massless theory given in \cite{Morita:2020liy}, which we matched with the boostrap in the main text.

In Fig.\,\ref{fig:e_two_matrix} of the main text we see that
for all couplings the $L=4$ bootstrap results lie within a narrow range bounded by (\ref{eq:ine}).

The expectation values $\langle \tr X^2 \rangle$ and $\langle \tr [X,Y]^2 \rangle$ in the trial wavefunction (\ref{eq:BOapp}) do not provide bounds in the way that the energy does, and therefore we have not included them in Fig. \ref{fig:e_two_matrix}. For completeness we note that these expectation values can be
computed from the minimizing numerical distribution $\rho(x)$ as
\begin{align}
\langle \tr X^2 \rangle & = \int dx \rho(x) x^2 \,, \\
\langle \tr [X,Y]^2 \rangle & = - \sum_{i, j = 1}^N \langle (x_i-x_j)^2 |y_{ij}|^2 \rangle = - \int \frac{dx dx' \rho(x)\rho(x') (x- x')^2 }{2\sqrt{m^2 + g^2 (x-x')^2}} \,, \\
\langle \tr Y^2 \rangle & = \sum_{i, j = 1}^N \langle |y_{ij}|^2 \rangle = \int \frac{dx dx' \rho(x)\rho(x')}{2\sqrt{m^2 + g^2 (x-x')^2}} \,.
\end{align}
The wavefunction (\ref{eq:BOapp}) is not rotationally symmetric and hence $\langle \tr X^2 \rangle \neq \langle \tr Y^2 \rangle$ in general.

\section{Numerical implementation}
\label{app:numeric}

In this section we provide more details about the bootstrap numerics. A Python implementation is available at \url{https://github.com/hanxzh94/matrix-bootstrap}. The variables under consideration are expectation values of single trace operators, with three types of constraints: linear, quadratic and semidefinite. In the following we discuss the representations of the variables and the constraints, some tricks in the implementation, and the non-convex optimization algorithm.

The variables to solve for are expectation values of single trace operators, which are represented as strings of matrices. Denote the set of all possible matrix symbols as $\mathcal{A}$, and strings of length $\leq L$, constructed from matrices in $\mathcal{A}$, as $\mathcal{S}_L$. For example, in the single matrix case, $\mathcal{A} = \{X, P\}$, $\mathcal{S}_2 = \{\emptyset, X, P, X X, XP, PX, PP\}$, where $\emptyset$ denotes the empty string. The corresponding expectation values are $\langle \tr I \rangle$, $\langle \tr X \rangle$, $\langle \tr P \rangle$, ..., $\langle \tr PP\rangle$. Note that the matrices are non-commutative quantum operators. The expectation values $v_i$ are then labeled by an index $i$, e.g., $v_0 = \langle \tr I \rangle = N$, $v_1 = \langle \tr X \rangle$, $v_2 = \langle \tr P \rangle$ and so on. Represented as matrices and vectors, the linear constraints can be written as $\sum_j M_{i j} v_j = 0$, the quadratic constraints $\sum_{j k} M_{i j k} v_j v_k + \sum_j N_{i j} v_j = 0$, and semidefinite constraints $\mathcal{M}_{i j} = v_{k_{i j}} \succeq 0$. In the semidefinite constraint each matrix entry $\mathcal{M}_{i j}$ is a single trace expectation value $v_{k_{i j}}$ at index $k_{i j}$, and $k_{i j}$ is a function of $i$ and $j$ to be discussed later.

Linear equalities come from symmetry, gauge and reality constraints. Symmetry constraints take the form of $\langle [H, \mathcal{O}] \rangle = 0$, where $H$ is the symmetry generator, and $\mathcal{O}$ is an arbitrary single trace operator in $\mathcal{S}_{2 L}$. If the commutator generates operators outside $\mathcal{S}_{2 L}$, the constraint is discarded. The quantum commutator of two single trace operators is also a single trace, so $\langle [H, \mathcal{O}] \rangle = 0$ is a linear equality of some single trace expectation values. Equation (\ref{eq:one_virial}) in the main text is an example. 

For gauge constraints $\langle \tr G \mathcal{O} \rangle = 0$ as in (\ref{eq:one_gcons}), both $G$ and $\mathcal{O}$ are matrices instead of trace operators. In this case $\mathcal{O}$ runs over strings in $\mathcal{S}_{2 L - 2}$, and $\tr G \mathcal{O}$ is a linear combination of single trace variables. For example, in the one matrix case, $G$ is given by (\ref{eq:GGG}). Then if we take $\mathcal{O} = XX$, the equality is 
\begin{equation}
    i \langle \tr X P X X \rangle  - i \langle \tr P X X X \rangle  + N \langle \tr X X \rangle = 0.
\end{equation}

The reality constraints are $\langle \mathcal{O}^\dagger \rangle - \langle \mathcal{O} \rangle^* = 0$, for $\mathcal{O}$ a single trace operator in $S_{2 L}$. If all matrices in $\mathcal{A}$ are Hermitian, $\mathcal{O}^\dagger$ is simply the reversed string of $\mathcal{O}$. The constraint then identifies two single trace expectation values. 

Quadratic constraints result from cyclicity of the trace. Classically $\tr A B = \tr B A$, but operators in $A$ and $B$ may not commute quantum mechanically. For any string in $\mathcal{S}_{2 L}$, we impose the equality from trying to move the first matrix in the trace to the last. Specifically, let the single trace operator be 
$
    A_{i_0 i_1} B^{(1)}_{i_1 i_2} \ldots B^{(r)}_{i_r i_0},
$
where $A, B^{(k)} \in \mathcal{A}$ and the repeated indices are summed over. The corresponding constraint is 
\begin{equation} \label{eq:cyc}
    A_{i_0 i_1} B^{(1)}_{i_1 i_2} \ldots B^{(r)}_{i_r i_0} -  B^{(1)}_{i_1 i_2} \ldots B^{(r)}_{i_r i_0} A_{i_0 i_1} = \sum_{k = 1}^r B^{(1)}_{i_1 i_2} \ldots [A_{i_0 i_1}, B^{(k)}_{i_k i_{k + 1}}] \ldots B^{(r)}_{i_r i_0},
\end{equation}
where the bracket is the quantum commutator. Assume that commutators of single matrices are
$
    [A_{i j}, B_{k l}] = c_{A B} \delta_{i l} \delta_{j k}
$
for some constant $c_{A B}$. The right hand side of (\ref{eq:cyc}) is then a sum of double trace operators
\begin{equation}
    \sum_{k = 1}^r c_{A B^{(k)}} \tr B^{(1)} \ldots B^{(k - 1)} \tr B^{(k + 1)} \ldots B^{(r)}.
\end{equation}
An explicit example is given in equation (\ref{eq:cycl}) of the main text. At large $N$ the expectation values of double trace operators factorize, so the left side of (\ref{eq:cyc}) is linear in expectation values $v_i$ and the other side is quadratic. These equalities are the quadratic relations $\sum_{j k} M_{i j k} v_j v_k + \sum_j N_{i j} v_j = 0$ mentioned previously. 

As discussed in the main text, positivity of certain operator expectation values requires that the matrix $\mathcal{M}_{i j} = \langle \tr \mathcal{O}_i^\dagger \mathcal{O}_j\rangle$ be positive semidefinite. Here $\mathcal{O}_i$ and $\mathcal{O}_j$ run over strings in $\mathcal{S}_L$, so that $\langle \tr \mathcal{O}_i^\dagger \mathcal{O}_j \rangle$ is an expectation value $v_{k_{i j}}$ in $\mathcal{S}_{2 L}$. The index $k_{i j}$, as a function of $i$ and $j$, is determined by the fact that the string $\mathcal{O}_{k_{i j}}$ is the string $\mathcal{O}_{i}^\dagger \mathcal{O}_j$.
In terms of the variables $v_i$, the positivity constraint is then that the matrix $\mathcal{M}_{i j} = v_{k_{i j}}$ should be positive semidefinite.

Before delivering the variables and the constraints to optimization, we discuss several implementation tricks used to simplify coding or improve computational efficiency. Firstly, all expectation values are scaled by proper factors of $N$ so that $N$ is not explicit in the numerics. The $N$ scaling can be determined from free theories and is $N^{l / 2 + 1}$ for a single trace operator with $l$ matrices. 

Secondly, some expectation values must vanish due to symmetries and hence are not included in the constraints. For one matrix quantum mechanics (\ref{eq:h}) expectation values of an odd number of matrices must vanish. For two matrix quantum mechanics (\ref{eq:h_two}) it is more efficient to work with the following matrix basis $\mathcal{A} = \{A, B, C, D\}$:
\begin{align} \label{eq:basis}
    A = P - i X - i (Q - i Y), \quad B = P + i X + i (Q + i Y), \nonumber\\
    C = P - i X + i (Q - i Y), \quad D = P + i X - i (Q + i Y).
\end{align}
The four matrices are eigenvectors of the $SO(2) \cong U(1)$ action with eigenvalues $-1, 1, 1, -1$. Hence $SO(2)$ rotation invariance is imposed if we only consider strings with $n(A) - n(B) - n(C) + n(D) = 0$, where, for example, $n(A)$ is the number of $A$'s in the string. The number of possible strings is thus significantly reduced.

Thirdly, for bosonic matrix models the wavefunction can be chosen as real, and hence expectation values of strings with an odd number of $P$'s (and an arbitrary number of $X$'s) must be purely imaginary, while strings with an even number of $P$'s must be real. This fact simplifies the reality constraints and reduces the number of real variables to optimize over. 

Lastly, the linear constraints $\sum_j M_{i j} v_j = 0$ can be solved to obtain a linearly independent set of variables $\widetilde{v}_i$, where $v_i = \sum_j K_{i j} \widetilde{v}_j$ and $\sum_j M_{i j} K_{j k} = 0$. Then the quadratic and semidefinite constraints are rewritten in terms of $\widetilde{v}_i$. The optimization is more efficient on this reduced set of variables.

In the optimization, the energy $\langle H \rangle$ is minimized subject to the constraints $\sum_j M_{i j} v_j = 0$, $\sum_{j k} M_{i j k} v_j v_k + \sum_j N_{i j} v_j = 0$ and $\mathcal{M}_{i j} \succeq 0$. The constraints are generally non-convex due to the presence of quadratic equalities. We employ a trust-region sequential semidefinite programming algorithm for the non-convex optimization \cite{nocedal2006numerical}. The algorithm iteratively searches for a local minimum of the goal function, and the basic idea is as follows. At each step, the quadratic constraint is approximated by its local linearization. With only linear and semidefinite constraints, the problem is convex and solved with semidefinite programming. The variables $v_i$ (or $\widetilde{v}_i$) are then updated with the solution of this local convex approximation, and the algorithm proceeds to the next step. Optimization finishes when the updates are smaller than some threshold. Expectation values of the energy and other trace operators at the local minimum are returned. 

\end{document}